\documentclass{PoS}
\usepackage{cite,url}
\usepackage[utf8]{inputenc}
\usepackage[LGR,T1]{fontenc}

\title{BSM physics at the LHeC and the FCC-he}

\ShortTitle{BSM physics at the LHeC and the FCC-he}

\author{Georges Azuelos\\
       Universit\'e de Montr\'eal, Montr\'eal,  Canada \\
       TRIUMF, Vancouver, Canada \\
        E-mail: \email{georges.azuelos@umontreal.ca}}

\author{Monica D'Onofrio\\
Department of Physics 
Oliver Lodge Laboratory,
University of Liverpool,
Oxford Street, Liverpool,
L69 7ZE\\
        E-mail: \email{Monica.D'Onofrio@cern.ch}}

\author{Oliver Fischer\\
        Institute for Nuclear Physics (IKP), Karlsruhe Institute of Technology, Hermann-von-Helmholtz-Platz 1, D-76344 Eggenstein-Leopoldshafen, Germany\\
        E-mail: \email{oliver.fischer@kit.edu}}

\author{\speaker{Jos\'e Zurita} \\%
  
        Institute for Nuclear Physics (IKP), Karlsruhe Institute of Technology, Hermann-von-Helmholtz-Platz 1, D-76344 Eggenstein-Leopoldshafen, Germany \\
Institute for Theoretical Particle Physics (TTP), Karlsruhe Institute of Technology, Engesserstrae 7, D-76128 Karlsruhe, Germany\\
        E-mail: \email{jose.zurita@kit.edu}}
        
\abstract{Electron-proton ($e^-p$) colliders are an ideal laboratory to study common features of electron and quarks with production via electroweak bosons, leptoquarks, multi-jet final states and very forward physics, due to their impressive pseudorapidity coverage.
  In addition to these physics cases, there exist a broad Beyond the Standard Model (BSM) program aimed at exploring the capabilities of the LHeC~\cite{AbelleiraFernandez:2012cc} and FCC-he~\cite{Zimmermann:2014qxa} for several New Physics scenarios.
Although their centre-of-mass energy is down with respect to a $pp$ collider by a factor of $\sqrt{E_p/E_e} \sim 10~(30)$ for the LHeC (FCC-he), 
they can be an invaluable tool to characterize BSM physics hints at $ee$ and $pp$ machines.

The aim of this talk is to provide, on behalf of the BSM $e^-p$ Working Group, an overview of the aforementioned BSM program, by briefly summarizing the existing studies and reporting on the most recent progress. We expect that the ample scope in terms of NP models to be tested would enhance the synergies between the BSM and  $e^-p$ communities.}

\FullConference{XXVI International Workshop on Deep-Inelastic Scattering and Related Subjects (DIS2018)\\
		16-20 April 2018\\
		Kobe, Japan}

\begin{document}

\section{Selected results}
An important number of phenomenological BSM studies for for $e^- p$ colliders have been carried out~\cite{Acar:2016rde,Han:2018rkz,Kuday:2013cxa,Wei:1900zz,Curtin:2017bxr,Duarte:2018xst,Antusch:2016ejd,Duarte:2014zea,Kuday:2017vsh,Cakir:2014swa,Azuelos:2017dqw,Sun:2017mue,Zarnecki:2008cp,Zarnecki:2001es,Zarnecki:2016akl,Liu:2017rjw,Lindner:2016lxq,Mondal:2015zba,Denizli:2017cfx,Wang:2017pdg,TurkCakir:2017rvu,Sarmiento-Alvarado:2014eha,Dutta:2013mva,Acar:2016rsw,Hernandez-Sanchez:2015bda,Das:2015kea,Sahin:2013vha,Ren-You:2014pqa,Ozansoy:2016ivj}. The most updated list of these studies can be found in~\cite{bsmatlhecweb}. The BSM scenarios under consideration include leptoquarks, $e-q$ compositeness, anomalous gauge bosons and top couplings~\footnote{Studies of anomalous Higgs couplings are part of the LHeC-FCC-eh Higgs working group.}. We stress that many of these results have been produced in the last year, which is a token of the attention that the BSM community is paying to the $e^- p$ colliders. Since it would be an impossible task to summarize all these studies in a few minutes, we will concentrate on a few examples, to give a taste of the $e^-p$ capabilities, and comparing the physics reach in these concrete scenarios between the LHeC and LHC, and also for the FCC-hh, FCC-he and FCC-ee.
\subsection{BSM Higgs sector}
\begin{figure}[tp]
\begin{center}
\includegraphics[height=4.4cm]{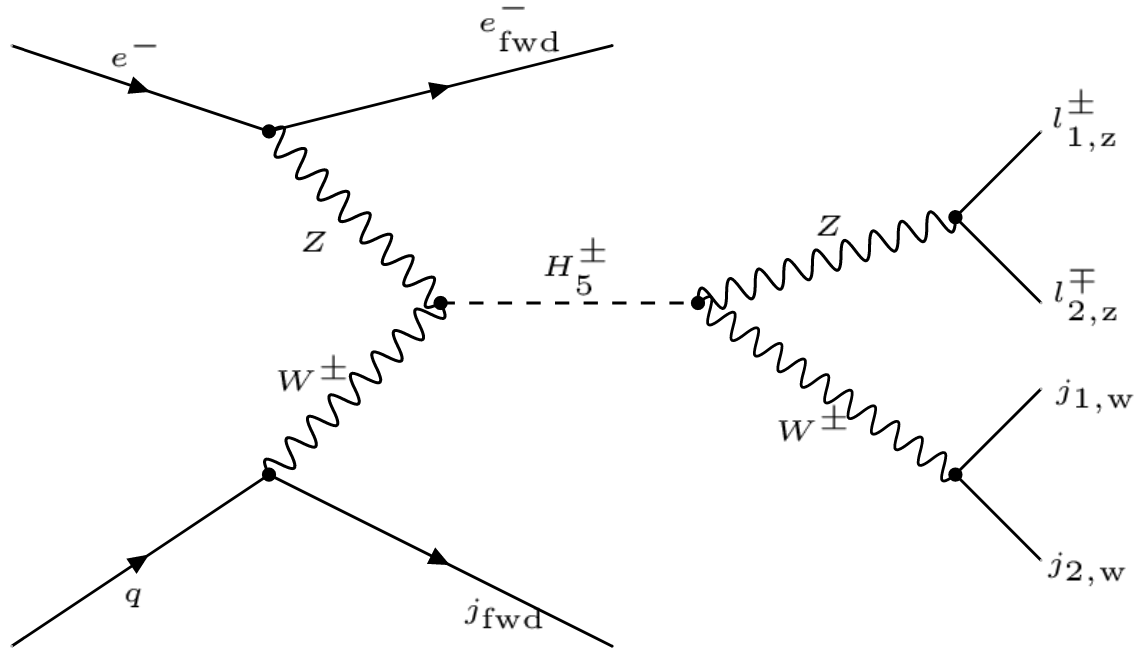}
\hspace{2mm}
\includegraphics[height=4.4cm]{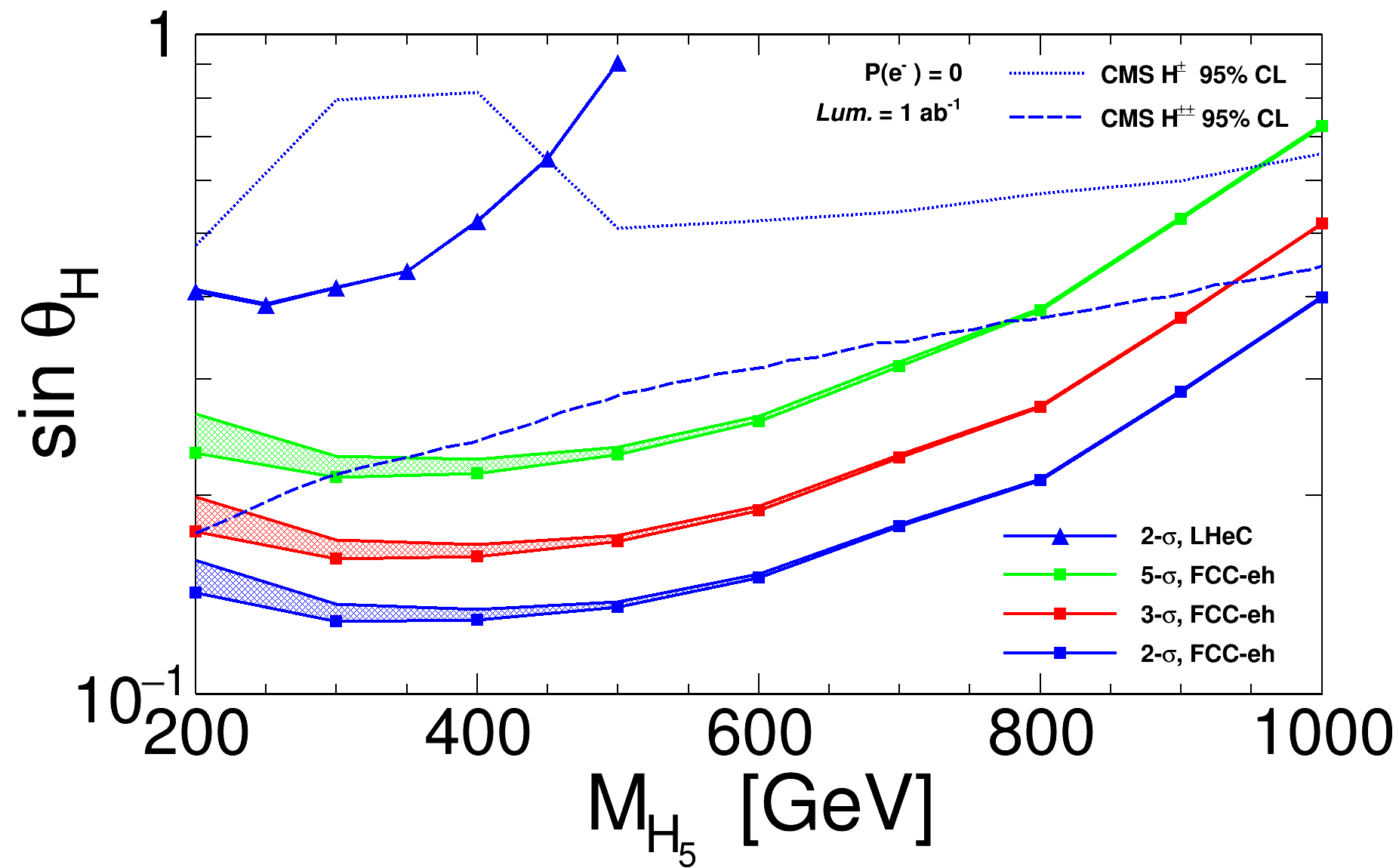}
\end{center}
%\vspace{-0.7cm} 
\caption{\label{fig:GMbounds}{ 
Left: Feynman diagram for charged Higgs production in the GM model at an $e^-p$ collider. Right: Expected bounds on charged Higgs from LHC, LHeC and FCC-eh. Taken from Ref.~\cite{Azuelos:2017dqw}  }}
\end{figure}

The Georgi-Machacek (GM) model is a triplet extension of the Higgs sector which preserves custodial symmetry.
The spectrum features a fermiophobic five-plet $(H_5^{\pm \pm},H_5^\pm,H_5^0)$ for which the phenomenology can be described
in terms of two parameters: $m_{H_5}$ and $\sin \theta_H$, which is the mixing angle in the neutral sector. In Fig.~\ref{fig:GMbounds} we see the main production mechanism at $e^-p$ colliders for the charged Higgs (left panel) and the expected CMS and $e^-p$ collider bounds on the charged states, showing that the LHeC would probe a region of parameter space beyond present CMS reach.
\subsection{SUSY (with or without R-parity)}
The searches for SUSY at pp colliders have two obvious loopholes where the mass reach is severely hindered. If R-parity is violated interactions of the form ${\cal L} \supset \lambda_{ijk} \hat{L}_i \hat{L}_j \hat{E}^C_{k} + \lambda^{\prime}  \hat{L}_i \hat{Q}_j \hat{D}^C_{k} $ are allowed (one SUSY particle in a vertex). The second one is the production of electroweak states (either promptly or long-lived), and in particular when the spectrum becomes compressed.

The RPV couplings can be probed by e.g: multi-lepton and multijet at the LHC. At the LHeC one can constrain anomalous $e$-$d$-$t$ interactions $\lambda^{\prime}_{131}$ < 0.03 and also the product $\lambda^{\prime}_{131} \lambda^{\prime}_{i33}$~\cite{Wei:1900zz}.~\footnote{The FCC-eh potential is under assessment by the authors of ~\cite{Wei:1900zz}. }.

\begin{figure}[tp]
\begin{center}
\includegraphics[height=4.1cm]{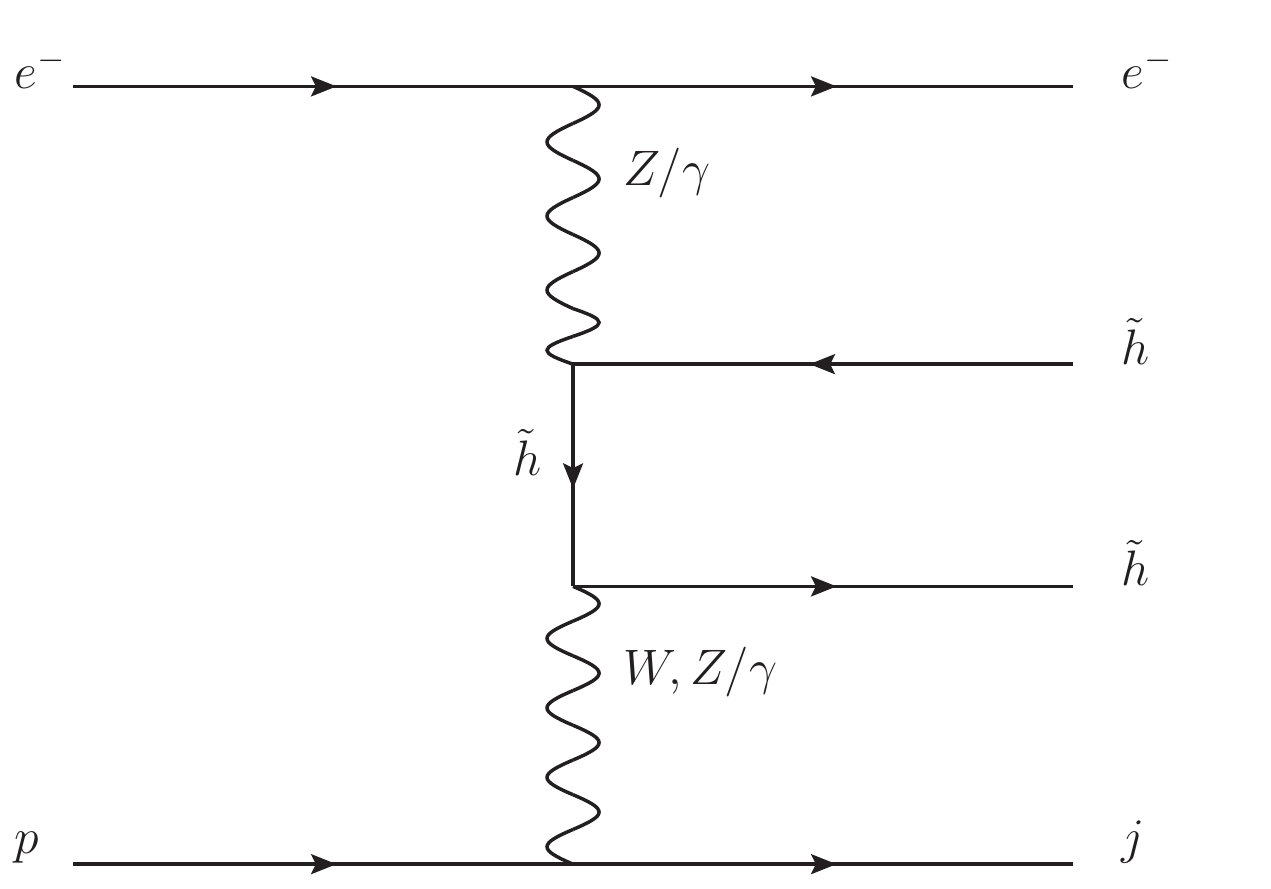}
\hspace{2mm}
\includegraphics[height=4.1cm]{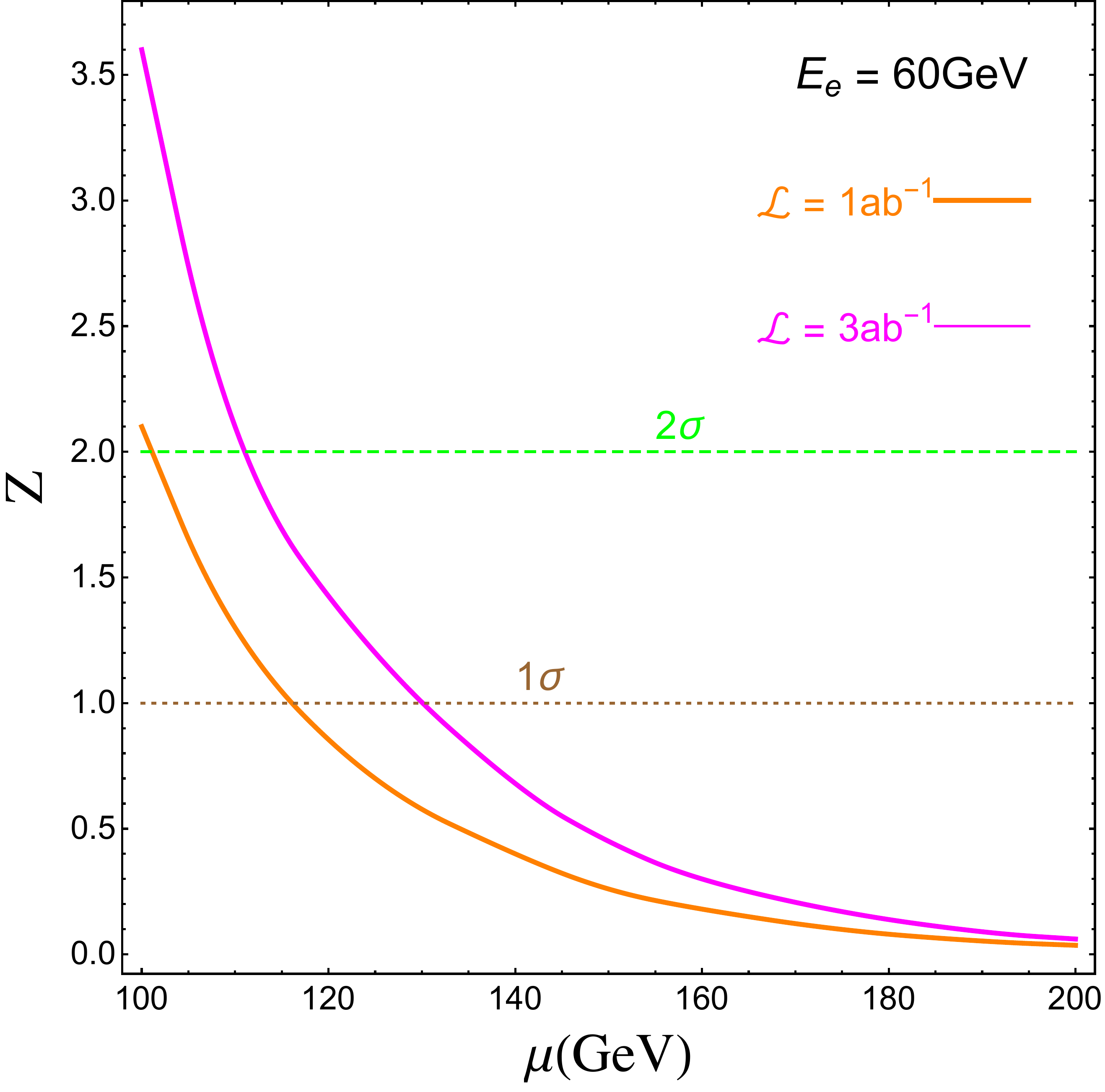}
\hspace{2mm}
\includegraphics[height=4.1cm]{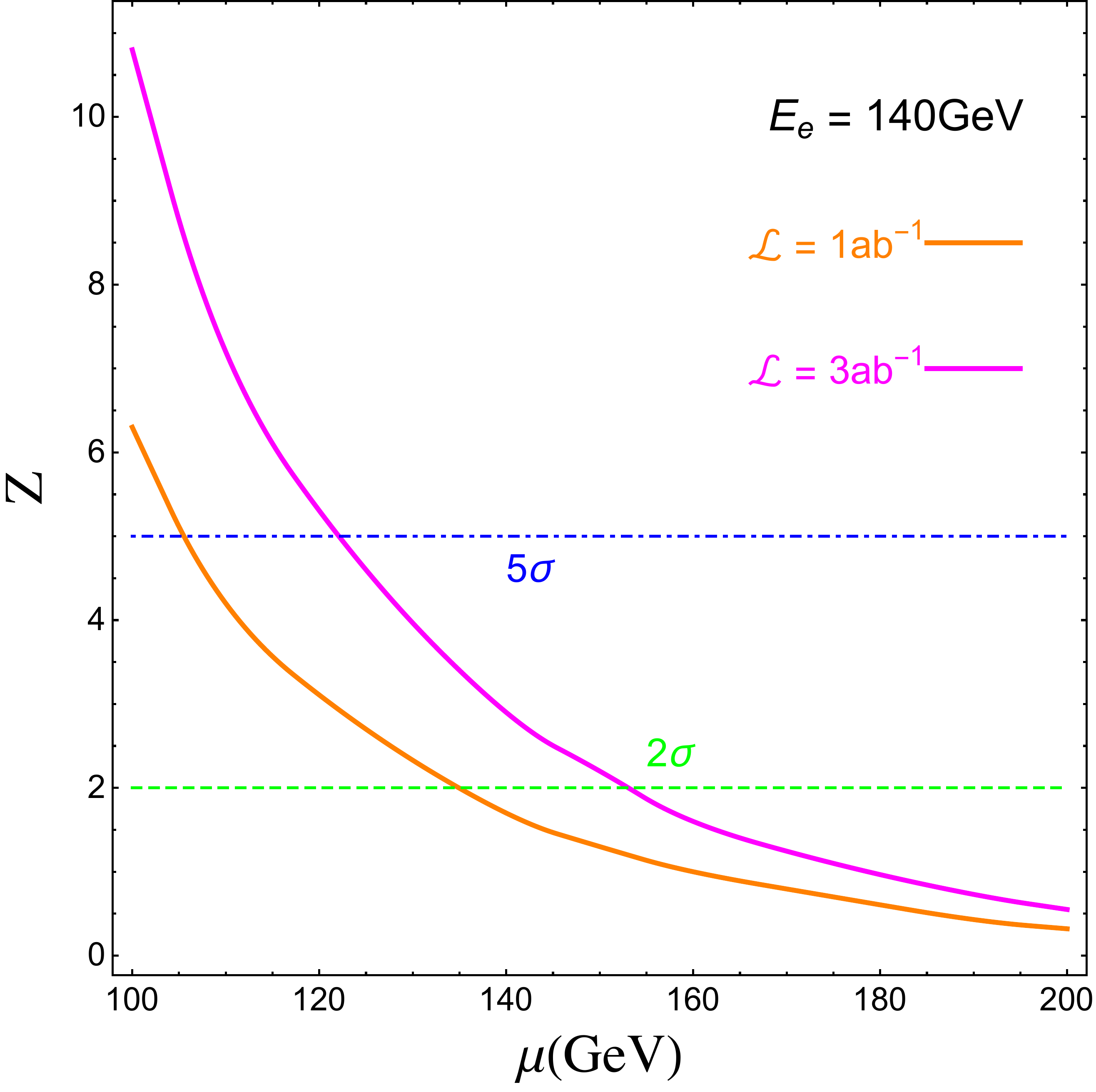}
\end{center}
%\vspace{-0.8cm} 
\caption{\label{fig:fcceh_higgsino}{ 
Feynman diagram for Higgsino pair production at an $e^-p$ collider (left). Expected LHeC (center) and FCC-eh (right) reach on Higgsino mass. Taken from ref.~\cite{Han:2018rkz} and ongoing work. }}
\end{figure}

Prompt Higgsino searches at the LHC in the compressed spectra can probe up to 150-250 GeV~\cite{Schwaller:2013baa,Barducci:2015ffa,Ismail:2016zby}, with the actual reach strongly depending on the level of systematic uncertainties. At a e-p collider these states are produced via VBF (fig.~\ref{fig:fcceh_higgsino}, left) and they can cover up to 150 GeV masses (fig.~\ref{fig:fcceh_higgsino}, right) if they decay promptly
 (for long-lived decays see next section).
\subsection{Long-lived particles}
\begin{figure}[!htp]
\begin{center}
\includegraphics[height=3.8cm]{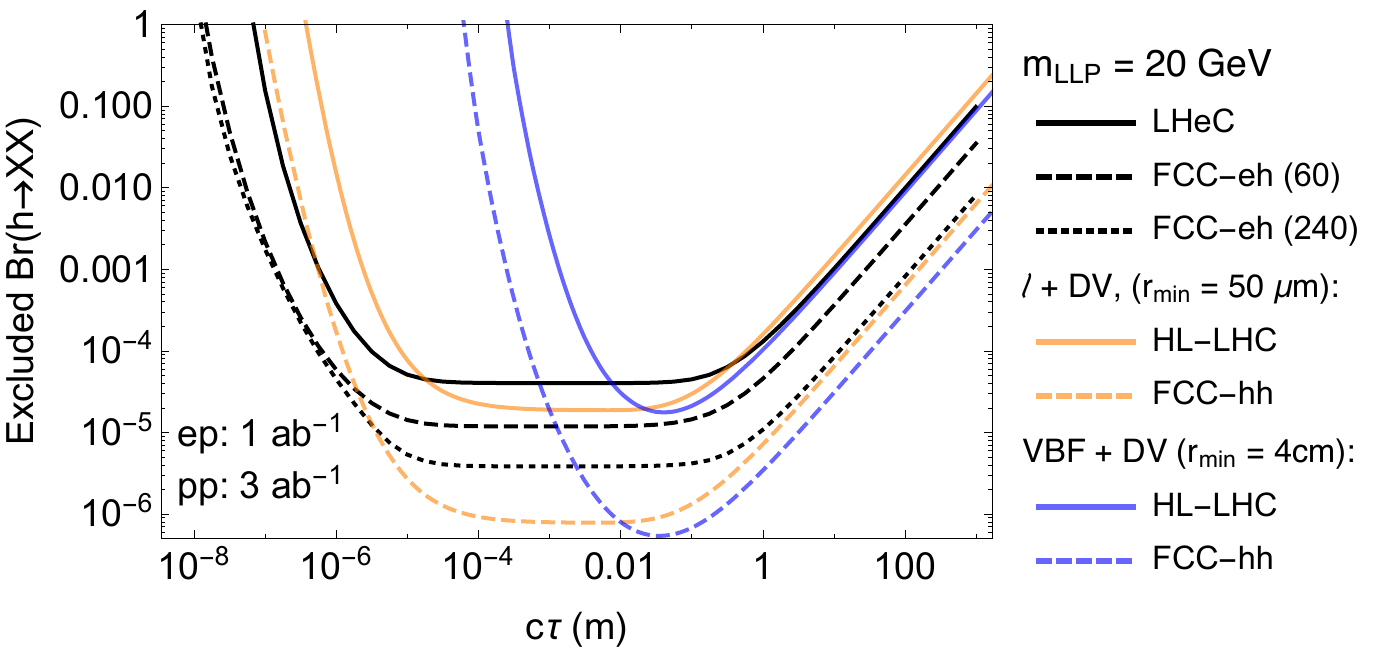}
\hspace{2mm}
\includegraphics[height=3.8cm] {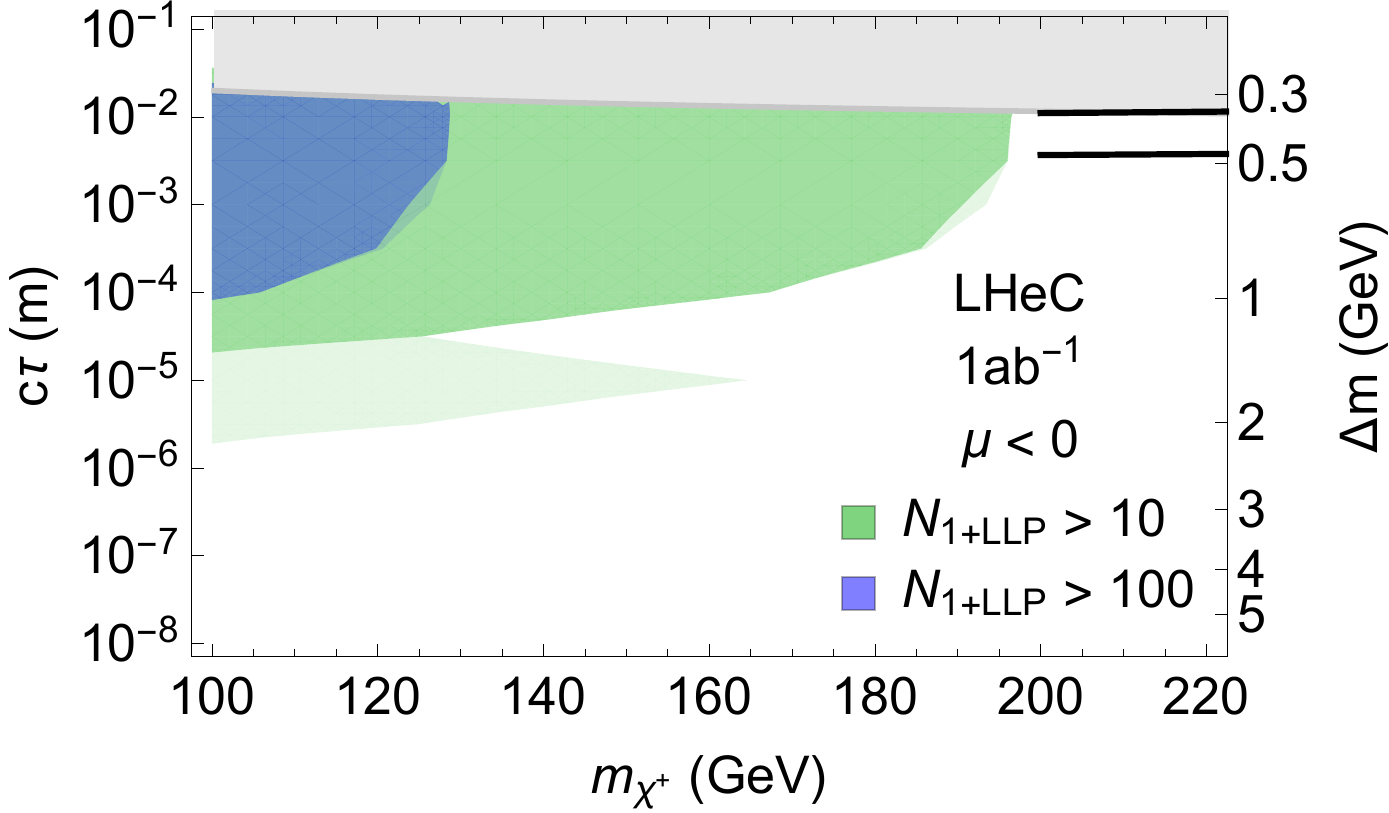}
\end{center}
%\vspace{-0.8cm} 
\caption{\label{fig:exothiggs}{Left: Reach for long-lived Higgsinos in the $m_{\chi}$-$c \tau$ plane, compared to disappearing tracks at the HL-LHC~\cite{Mahbubani:2017gjh}. Right: Exotic Higgs branching fraction that can be excluded by $pp$ and $e$-$p$ colliders~\cite{Curtin:2017bxr}. }}
\end{figure}

Long-lived-particles (LLPs) are a well motivated physics scenario, as many theories aiming at solving fundamental problems of the SM naturally include them. They lead to spectacular new physics signals due to the conspicuous nature of the sought final states. Regarding the LHeC, studies have been performed for non-prompt Higgsino decays in SUSY scenarios with a compressed spectrum and exotics Higgs decays into a pair of light LLP's. For the former we present the LHeC reach in the left panel of figure~\ref{fig:exothiggs}~\cite{Curtin:2017bxr,Curtin:2018qot} and for the latter we show the exotic Higgs branching fraction that several $pp$ and $e^-p$ colliders can test in figure~\ref{fig:exothiggs}. 
\subsection{Sterile neutrinos}
The presence of (feeble-interacting) sterile neutrinos can explain the generation of neutrino masses via a lowscale seesaw mechanism. Currently the mixing angle with the electrons is constrained to $\theta_e \lesssim 10^{-3}$. Their collider phenomenology is dominated by lepton-flavor violating processes and by displaced vertices for masses below $m_W$. For further details see~\cite{Antusch:2016ejd}.

\begin{figure}[tp]
\begin{center}
\includegraphics[height=3.3cm]{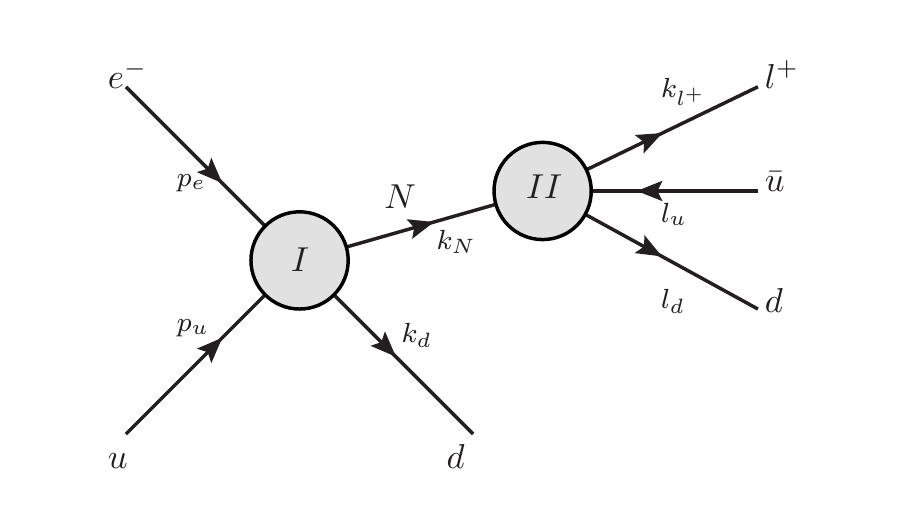}
\hspace{1mm}
\includegraphics[height=3.3cm]{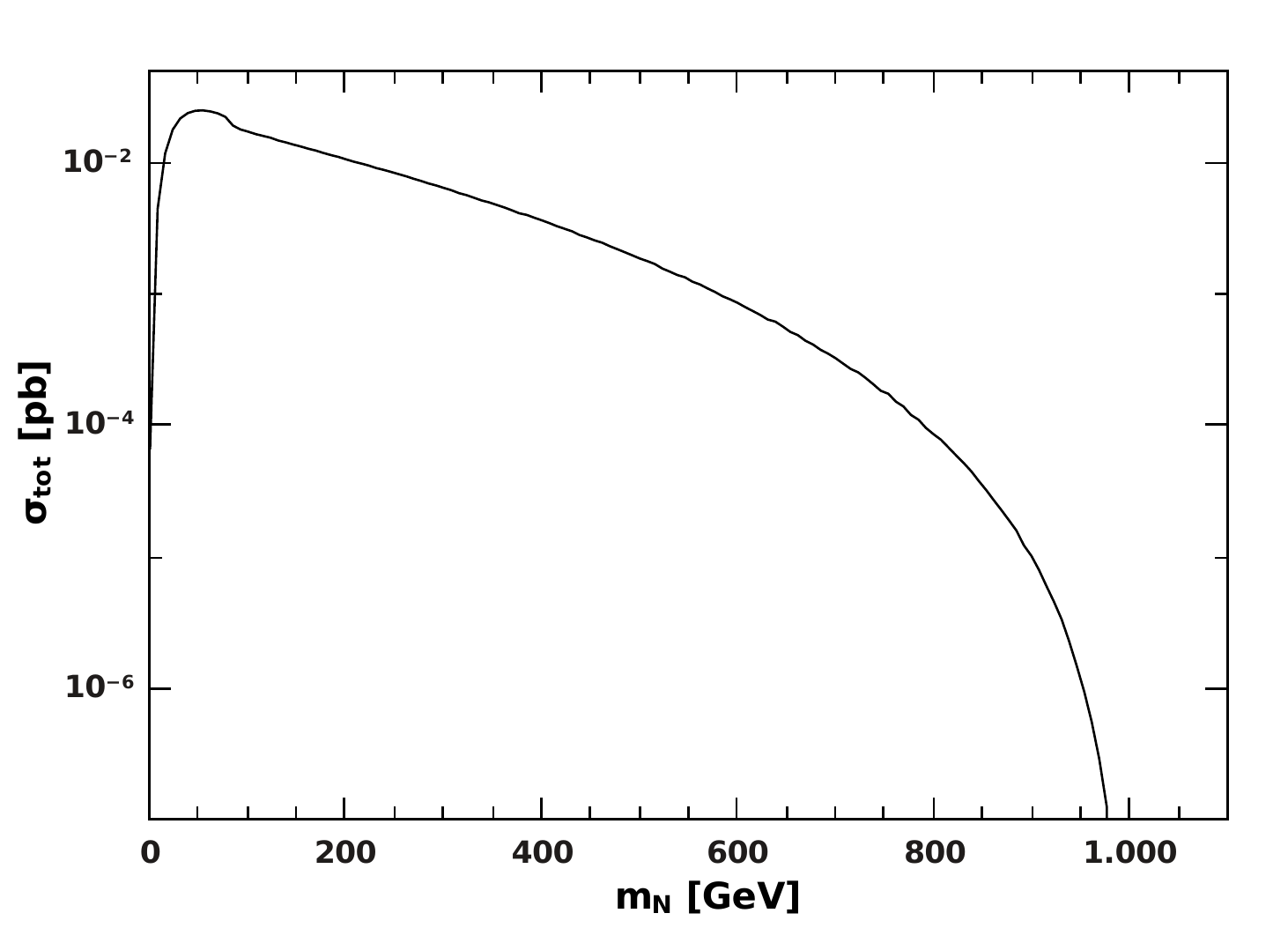}
\hspace{1mm}
\includegraphics[height=3.3cm]{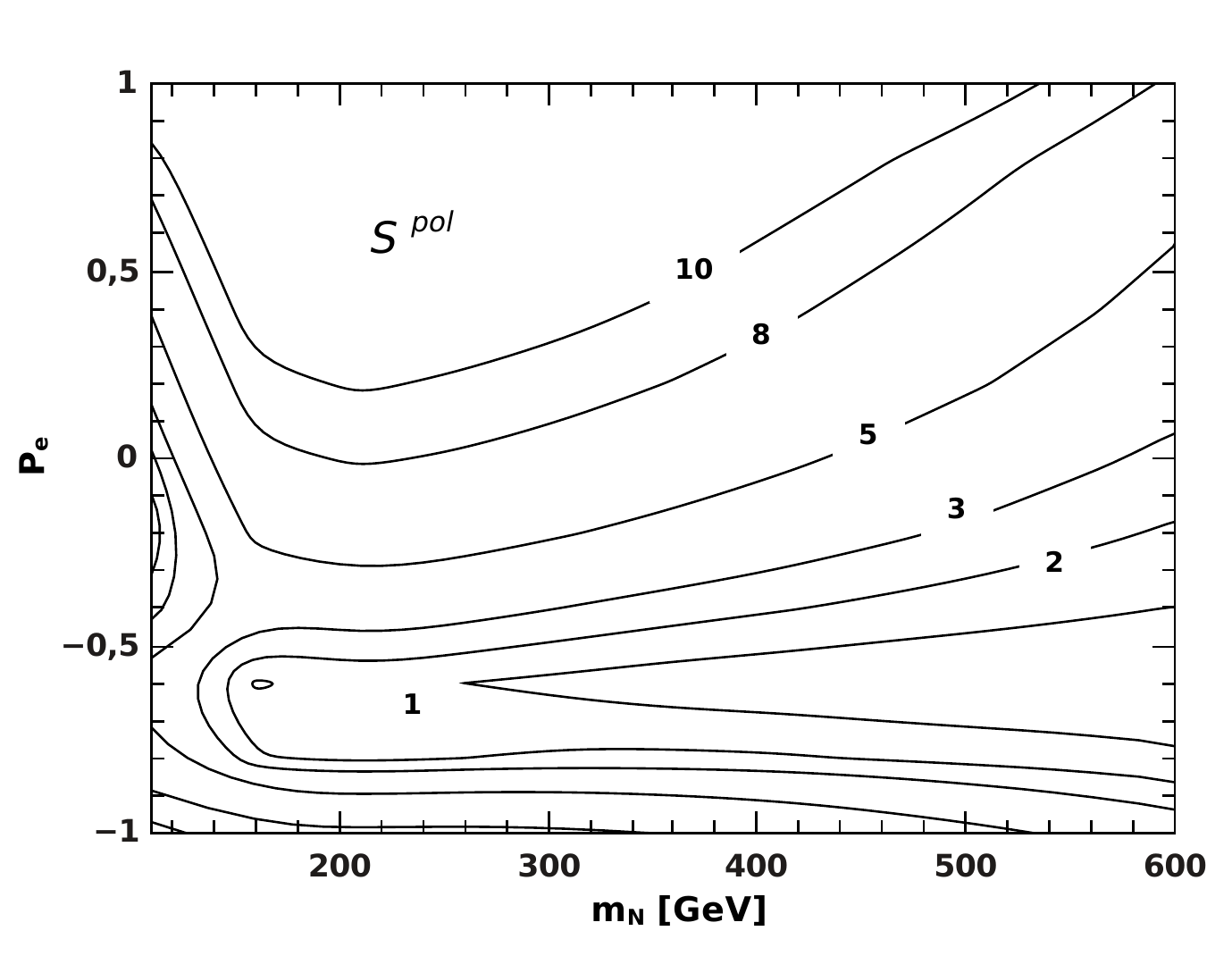}
\end{center}
%\vspace{-0.7cm} 
\caption{\label{fig:HMN}{ 
Feynman diagram for Majorana production at an $e^-p$ collider (left). Taken from ref.~\cite{Duarte:2018xst}. }}
\end{figure}

In this context, ref.~\cite{Duarte:2018xst} studied the production of heavy Majorana neutrinos via new operators which is depicted in the left panel of fig~\ref{fig:HMN}, and the corresponding LHeC cross section is shown in the middle panel. These operators lead to a non-trivial Lorentz structure of the production process, which can be probed and studied via polarized electron beams (right panel of fig~\ref{fig:HMN}) , and the $e^-p$ collider has the unique capability to disentangle the vector and scalar operators.
\section{Conclusions}
We have seen with several examples how an $e^-p$ collider nicely complements the physics program of their $pp$ and $e^+ e^-$ cousins, offering rich opportunities for a plethora of BSM scenarios. They are ideal to study the properties of new particles that couple to electrons and quarks, but they also provide great coverage in previously unexplored signals, like an unprecedented reach for low lifetimes for signals buried in the $pp$ hadronic noise. It is thus crucial to carry out a wide program of preliminary studies to fully determine the New Physics reach capabilities of the future $e^- p$ facilities, beyond their already guaranteed return in terms of understanding the proton structure.

\end{document}